\magnification 1200
\centerline {\bf  Quantum Stochastic Models with Hydrodynamical Behaviour}
\vskip 0.5cm
\centerline {\bf by Geoffrey Sewell}
\vskip 0.3cm
\centerline {\bf Department of Physics, Queen Mary, University of London,}
\vskip 0.2cm
\centerline  {\bf Mile End Road, London E1 4NS}
\vskip 0.2cm
\centerline {(e-mail: g.l.sewell@qmul.ac.uk)}
\vskip 1cm
\centerline {\bf Abstract}
\vskip 0.3cm
We construct a class of  quantum stochastic models of reservoir driven many-particle 
systems that are the natural counterparts of certain extensively studied classical ones, 
which have been shown to exhibit good hydrodynamical behaviour. Our treatment of 
these models achieves two main aims. The first is to show that they enjoy the 
hydrodynamical properties of their classical counterparts. The second is to show that they 
satisfy the key assumptions of the general quantum macrostatistical scheme, presented in 
earlier works by the author, which served to expose certain generic large scale features of  
nonequilibrium steady states, e.g. the long range hydrodynamical correlations that they 
carry. In this way we establish the viability of that scheme. 
\vfill\eject
\centerline {\bf I. Introduction} 
\vskip 0.3cm
The purpose of this note is to bring together two developments in the theory of  the 
relationship between the hydrodynamical and the microscopic pictures of  reservoir 
driven macroscopic systems.. The first of these developments, which we shall refer to as 
(I), is a body of work concerned with the derivation of hydrodynamics from the 
microscopic dynamics of  a class of classical stochastic models [1-5]. Most interestingly, 
the hydrodynamical fluctuations of these models about their nonequilibrium steady states 
have been shown to carry long range spatial correlations [3-5] and to conform to a 
generalised version [5] of the Onsager-Machlup process [6]. The second development, 
which we shall refer to as (II),  is a general, model-independent, quantum macrostatistical 
treatment [7,8] of hydrodynamical fluctuations about nonequilibrium steady states, that is 
based on certain hypotheses of chaoticity and local equilibrium, together with a 
generalised version of  Onsager\rq s regression hypothesis [9]. On this very general basis, 
we have shown that, as in the special classical models of [3-5], the hydrodynamical 
fluctuations about nonequilibrium steady states carry long range spatial correlations and 
execute a generalised Onsager-Machlup process.
\vskip 0.2cm
Our objectives here are to extend the constructive developments of (I) to the quantum 
regime and to show that the resultant models satisfy the assumptions of the general 
quantum macrostatistical theory of (II). The latter objective is thus designed to show that 
the  \lq axiomatic\rq\ scheme of (II) is viable.
\vskip 0.2cm
Our approach to these objectives is based on a construction whereby we extend the 
generic classical stochastic model, ${\Sigma}_{cl}$, of  (I) to a quantum system, 
${\Sigma}$, in such a way that 
\vskip 0.2cm\noindent
(i) the abelian algebra of observables, ${\cal B}$, of ${\Sigma}_{cl}$ is a subalgebra of 
the nonabelian one, ${\cal A}$, of ${\Sigma}$;
\vskip 0.2cm\noindent
(ii) ${\cal B}$ is stable under the dynamics of  ${\Sigma}$;
\vskip 0.2cm\noindent
(iii) the nonequilibrium steady state of ${\Sigma}_{cl}$ is just the restriction to 
${\cal B}$ of that of ${\Sigma}$; and
\vskip 0.2cm\noindent
(iv) the hydrodynamical observables of ${\Sigma}$ are precisely those of 
${\Sigma}_{cl}$,
\vskip 0.2cm\noindent
Thus, the construction of the quantum model ${\Sigma}$ in this way permits us to 
exploit some of the powerful results obtained for the hydrodynamical properties of its 
classical counterpart, ${\Sigma}_{cl}$. In particular, it enables us to verify that 
${\Sigma}$ enjoys the hydrodynamical properties of ${\Sigma}_{cl}$ and, moreover, 
that it satisfies the basic assumptions of (II). This latter result therefore establishes that 
the scheme (II) is viable. 
\vskip 0.2cm
We present the formulation of the models ${\Sigma}_{cl}$ and ${\Sigma}$ in Section 2, 
and establish there the above properties (i)-(iv). Thus the hydrodynamical picture of 
${\Sigma}$ reduces to that of ${\Sigma}_{cl}$. We formulate this picture explicitly in 
Section 3 and extend it to fluctuations about a nonequilibrium steady state in Section 4, 
where we specify the regression, chaoticity and local equilibrium hypotheses on which 
the theory of (II) was based. In Section 5 we prove the validity of these hypotheses for 
the present  stochastic model and thereby establish the viability of the scheme of [II]. We 
conclude in Section 6 with a brief comment about an outstanding problem in the theory 
of quantum stochastic processes. 
\vskip 0.5cm
\centerline {\bf  2. The Classical and Quantum Stochastic Models} 
\vskip 0.3cm
The generic model with which we shall be concerned, whether classical or quantum, is a 
system of $N$ identical particles that live in a bounded region ${\Omega}_{N}$ of the 
$d$-dimensional lattice ${\bf Z}^{d}$ and are coupled to reservoirs at its boundary. 
 More specifically, ${\Omega}_{N}$ is assumed to be the subset of  ${\bf Z}^{d}$ 
contained within the dilation by a certain factor,  $L_{N}$, of a fixed, $N$-independent 
bounded open connected  region ${\Omega}$ of the Euclidean space ${\bf R}^{d}$. 
Thus, ${\Omega}_{N}={\bf Z}^{d}{\cap}(L_{N}{\Omega})$. We define its boundary, 
${\partial}{\Omega}_{N}$,  to comprise the sites in ${\Omega}_{N}$ with at least one 
nearest neighbour that lies outside that region, and we define 
${\rm Int}({\Omega}_{N})$, the interior of ${\Omega}_{N}$, to be 
${\Omega}_{N}{\backslash}{\partial}{\Omega}_{N}$. This latter region thus consists 
of the sites in ${\Omega}_{N}$ whose nearest neighbours also lie in ${\Omega}_{N}$.
\vskip 0.2cm  
We assume that the volume of ${\Omega}$ is unity, that its boundary, 
${\partial}{\Omega}$, is smooth and that the mean particle number density, ${\nu}$, of 
the system is $N$-independent. Thus
$$L_{N}=(N/{\nu})^{1/d}.\eqno(2.1)$$
We assume that the dynamics of the model corresponds to a stochastic process whereby 
the particles jump between nearest neighbouring lattice sites according to probabilistic 
laws that will be prescribed below. 
\vskip 0.3cm
\centerline {\bf 2.1. The Classical Model.}
\vskip 0.3cm
For the classical model, ${\Sigma}_{cl}$, we denote by $n_{x}$ the number of particles 
at the site $x$. In the case where an exclusion principle is operative, $n_{x}$ is restricted 
to the values $0$ and $1$: otherwise it may take any non-negative integral value. Thus a 
particle configuration is a map $n:x{\rightarrow}n_{x}$ of ${\Omega}_{N}$ into a set 
$K$, which is either ${\lbrace}0,1{\rbrace}$ or ${\bf N}$ according to whether or not an 
exclusion principle is operative. We take the algebra, ${\cal B}$, of bounded observables 
of the system to comprise the bounded, complex valued functions on the configuration 
space ${\Gamma}=K^{{\Omega}_{N}}$, with supremum norm. Thus, equipping 
${\Gamma}$ with the discrete topology, ${\cal B}={\cal C}({\Gamma})$, the 
$C^{\star}$-algebra of bounded continuous functions on ${\Gamma}$. For 
$x,y{\in}{\Omega}_{N}$, we define $n{\rightarrow}n^{x,y}$ to be the transformation 
of ${\Gamma}$  corresponding to the transfer of a particle from $x$ to $y$, provided 
that that transfer is kinematically admissible, i.e. that $(n_{x}-1)$ and $(n_{y}+1)$ lie in 
$K$: otherwise we define $n^{x,y}$ to be simply $n$. Likewise we define 
$n^{x,{\pm}}$ to be the modifications of $n$ corresponding to increments ${\pm}1$ in 
$n_{x}$, provided that $(n_{x}{\pm}1){\in}K$: otherwise we define $n^{x,{\pm}}=n$. 
We assume that the dynamics of the system is given by a continuous one-parameter 
semigroup,  
${\phi}_{cl}({\bf R}_{+})={\lbrace}{\phi}_{cl}(t){\vert}t{\in}{\bf R}_{+}{\rbrace}$ 
of linear, positivity preserving transformations of ${\cal B}$. We denote its generator by 
${\cal G}_{cl}$ and we shall presently specify its explicit form for two models, namely 
those conventionally termed [1-5] the simple exclusion model and the zero range model.
\vskip 0.3cm
{\it The Simple Exclusion Model.} . For this model,  $K={\lbrace}0,1{\rbrace}$ and 
${\cal G}_{cl}$ takes the following form
$${\cal G}_{cl}f(n)=
{\sum}_{x,y{\in}{\Omega}_{N}}^{\prime}n_{x}
(1-n_{y})[f(n^{x,y})-f(n)]+$$
$${\sum}_{b{\in}{\partial}{\Omega}_{N}}r_{b}n_{b}
\bigl(f(n_{b}^{-})-f(n)\bigr)+ {\sum}_{b{\in}{\partial}{\Omega}_{N}}
h(b/L_{N})(1-n_{b})\bigl(f(n_{b}^{+})-f(n)\bigr),\eqno(2.2)$$
where the prime over the first sum signifies that summation is confined to nearest 
neighbours,  $h$ is a smooth positive-valued function on ${\partial}{\Omega}$ and 
$r_{b}$ is the number of nearest neighbours of $b({\in}{\partial}{\Omega}_{N})$ on 
the lattice ${\bf Z}^{d}$ that lie outside ${\Omega}_{N}$. Thus, the first sum 
represents the jumps between nearest neighbouring sites of the particles in the interior of 
${\Omega}_{N}$, the second the escape of particles across its boundary and the third the 
supply of particles by external sources at the boundary.  
\vskip 0.3cm
{\it The Zero Range Model.} For this model, $K={\bf N}$ and, in the same notation as in 
Eq. (2.2),  ${\cal G}_{cl}$ takes the following form.
$${\cal G}_{cl}f(n)=
{\sum}_{x,y{\in}{\Omega}_{N}}^{\prime}g(n_{x})\bigl(f(n^{x,y})-f(n)\bigr)+$$
$${\sum}_{b{\in}{\partial}{\Omega}_{N}}r_{b}g(n_{b})\bigl(f(n_{b}^{-})-
f(n)\bigr)+{\sum}_{b{\in}{\partial}{\Omega}_{N}}
h(b/L_{N})\bigl(f(n_{b}^{+})-f(n_{b})\bigr),\eqno(2.3)$$
where $g$ is a positive valued, non-increasing function on $K$ for which $g(0)=0$ and 
${\rm sup}_{k}\bigl(g(k+1)-g(k)\bigr)$ is finite.
\vskip 0.3cm
{\it Note.} It follows from Eqs. (2.2) and (2.3) that in the cases of the simple exclusion 
and the zero range models ${\cal G}n_{x}$ takes the forms $({\Delta}n)_{x}$ for the 
and $\bigl({\Delta}(g{\circ}n)\bigr)_{x}$, respectively, for $x{\in}
{\rm Int}({\Omega}_{N})$, where ${\Delta}$ is the discrete Laplacian defined by the 
formula
$$({\Delta}f)_{x}={\sum}_{y{\in}{\Omega}_{N}}^{\prime}(f_{y}-f_{x}),$$
and the prime over ${\Sigma}$ again indicates that the sum is taken over sites $y$ that 
are the nearest neighbours of $x$.  Hence, for both models, the dynamics of the field $n$ 
is diffusive.
\vskip 0.5cm 
\centerline {\bf 2.2. The Quantum Model}
\vskip 0.3cm
We take the quantum model ${\Sigma}$ to be a system of fermions or bosons according 
to whether or not the exclusion principle is operative. In either case we formulate the 
model in a standard way in terms of the Fock space ${\cal H}$ and the creation and 
destruction operators 
${\lbrace}a_{x}^{\star},a_{x}{\vert}x{\in}{\Omega}_{L}{\rbrace}$ that act therein 
according to the following defining conditions.
\vskip 0.2cm\noindent
(a)  ${\cal H}$ contains a vector ${\Phi}$ that is annihilated by the action of each of the 
$a_{x}$\rq s and is cyclic with respect to the polynomials in the $a_{x}^{\star}$\rq s; 
and 
\vskip 0.2cm\noindent
(b) the operators $a_{x}$ and $a_{x}^{\star}$ satisfy the canonical commutation or 
anticommutation relations, namely
$$[a_{x},a_{y}^{\star}]_{\mp}={\delta}_{x,y}I; \ [a_{x},a_{y}]_{\mp}=0 \ {\forall} \ 
x,y{\in}{\Omega}_{N}, \eqno(2.4)$$
according to whether the system consists of bosons or fermions. For either case, we 
define the number operator
$${\hat n}_{x}=a_{x}^{\star}a_{x} \ {\forall} \ x{\in}{\Omega}_{N}.\eqno(2.5)$$
It follows immediately from Eqs. (2.4) and (2.5) that the ${\hat n}_{x}$\rq s 
intercommute and thus constitute a classical field ${\hat n}:=
{\lbrace}{\hat n}_{x}{\vert}x{\in}{\Omega}_{N}{\rbrace}$. We denote by ${\psi}(n)$ 
the simultaneous eigenvector of these operators ${\hat n}_{x}$ with corresponding 
eigenvalues $n_{x}$, i.e.
$${\hat n}_{x}{\psi}(n)=n_{x}{\psi}(n) \ {\forall} \ x{\in}{\Omega}_{N}.\eqno(2.6)$$
It follows from this formula and our specifications of ${\cal H}$ that the vectors 
${\psi}(n)$ form a complete orthogonal basis for this space as $n$  runs through the 
classical configuration space ${\Gamma}=K^{{\Omega}_{N}}$, with $K={\bf N}$ or 
${\lbrace}0,1{\rbrace}$  according to whether the particles are bosons or fermions. 
Further, by Eqs. (2.4)-(2.6),
$$a_{x}{\psi}(n)=n_{x}^{1/2}{\psi}(n^{x,-}) \ 
{\rm and} \ a_{x}^{\star}{\psi}(n)=
(1{\pm}n_{x})^{1/2}{\psi}(n^{x,+})\eqno(2.7)$$
and hence 
$$a_{y}^{\star}a_{x}{\psi}(n)=
\bigl(n_{x}(1{\pm}n_{y})\bigr)^{1/2}{\psi}(n^{x,y}),\eqno(2.8)$$
where $n^{x,y}$ and $n^{x,{\pm}}$ are defined as in Section 2.1 and ${\pm}$ signifies 
the boson-fermion alternatives. 
\vskip 0.2cm
We denote by ${\cal F}$ the additive group of bounded continuous real-valued functions 
${\theta}:x{\rightarrow}{\theta}_{x}$ on ${\Omega}_{N}$ and we define the unitary 
representation $U$ of ${\cal F}$  by the formula 
$$U({\theta})=
{\rm exp}\bigl(i{\sum}_{x{\in}{\Omega}_{L}}{\theta}_{x}{\hat n}_{x}\bigr),
\eqno(2.9)$$
\vskip 0.2cm
We take the algebra, ${\cal A}$, of bounded observables of ${\Sigma}$ to be that of the 
bounded operators in ${\cal H}$ and we define ${\gamma}$ to be the representation of 
${\cal F}$ implemented by $U$ in ${\rm Aut}({\cal A})$, i.e.
$${\gamma}({\theta})A=U({\theta})AU({\theta})^{-1} \ {\forall} \ A{\in}{\cal A}.
\eqno(2.10)$$
Thus, by Eqs. (2.4)-(2.6), (2,9) and (2.10), ${\gamma}({\theta})$ is the local gauge 
automorphism given by the formula 
$${\gamma}({\theta})a_{x}=a_{x}{\exp}(-i{\theta}_{x})\eqno(2.11)$$
\vskip 0.2cm
We define ${\cal B}$ to be the locally gauge invariant subalgebra of ${\cal A}$  i.e., by 
Eq. (2.10), the set of elements of ${\cal A}$ that commute with all the $U({\theta})$\rq s. 
It follows from this definition and Eq. (2.9) that ${\cal B}$ comprises the elements $B$ 
of  ${\cal A}$ for which
$$U({\theta})B{\psi}(n)=BU({\theta}){\psi}(n)=
{\rm exp}\bigl(i{\sum}_{x{\in}{\Omega}_{N}}
n_{x}{\theta}_{x}\bigr)B{\psi}(n) \ {\forall} \ n
{\in}{\Gamma}, \ {\theta}{\in}{\cal F},$$
i.e. for which $B{\psi}(n)$ is a simultaneous eigenvector of the ${\hat n}_{x}$\rq s with 
corresponding eigenvalues $n_{x}$. This signifies that $B{\psi}(n)=F(n){\phi}(n)$, 
where $F$ is  some bounded complex-valued function on ${\Gamma}$, i.e. that $B$ is 
the operator $F({\hat n})$, defined by the formula  
$$F({\hat n}){\psi}(n)=F(n){\psi}(n) \ {\forall} \ n{\in}{\Gamma}.\eqno(2.12)$$
Thus we have established the following proposition.
\vskip 0.3cm
{\bf Proposition 2.1.} {\it (1) ${\cal B}$ comprises the functions of ${\hat n}$; and 
\vskip 0.2cm\noindent
(2) the mapping $F({\hat n}){\rightarrow}F(n)$ of ${\cal B}$ onto 
${\cal C}({\Gamma})$, the algebra of observables of  ${\Sigma}_{cl}$, is a 
$C^{\star}$-isomorphism. Hence ${\cal B}$ may be identified with  the latter algebra.} 
\vskip 0.2cm
We assume that the dynamics of the model ${\Sigma}$ is given by a strongly continuous 
one-parameter semigroup ${\phi}({\bf R}_{+}) ={\lbrace}{\phi}(t){\vert}t{\in}
{\bf R}_{+}{\rbrace}$ of completely positive identity preserving contractions of  
${\cal A}$. Its generator ${\cal G}$ therefore takes the standard form for that of a 
quantum dynamical semigroup, namely [10, 11]
$${\cal G}A=i[H,A]_{-}+{\sum}_{j}\bigl(V_{j}^{\star}AV_{j}-
{1\over 2}[V_{j}^{\star}V_{j},A]_{+}\bigr) \ {\forall} \ A{\in}{\cal A},\eqno(2.13)$$  
where $H$ is a self-adjoint element of ${\cal A}$ and $V_{j}$ and 
${\sum}_{j}V_{j}^{\star}V_{j}$ also belong to this algebra. Thus,  the dynamics of the 
model is determined by $H$ and the $V$\rq s. We shall now specify thses operators for  
quantum versions of the simple exclusion and zero range models.
\vskip 0.3cm
{\it The Quantum Simple Exclusion Model.} In view of the requirement of an exclusion 
principle, we take the particles of this model to be fermions. We construct its dynamical 
semigroup ${\phi}({\bf R}_{+})$ in such a way as to obtain a natural correspondence 
between its generator and that of  ${\phi}_{cl}({\bf R}_{+})$, as given by  Eq. (2.2). 
Specifically we choose the operators $H$ and the $V$\rq s of Eq. (2.13) in the following 
way. 
\vskip 0.2cm\noindent
(a) We take $H$ to be zero, since ${\cal G}_{cl}$ contains no Hamiltonian part.
\vskip 0.2cm\noindent
(b) Since the indices involved in the structure of ${\cal G}_{cl}$ comprise the nearest 
neighbouring pairs $(x,y)$ of sites of ${\Omega}_{N}$ together with the boundary sites 
$b$ of ${\partial}{\Omega}_{N}$,  we assume that the index $j$ of Eq. (2.13) also runs 
through just these sets. 
\vskip 0.2cm\noindent
(c) For $j=(x,y)$,  we choose $V_{j}$ to be $a_{y}^{\star}a_{x}$, since the first 
summand of Eq. (2.13) then represents the transfer of a particle from $x$ to $y$, with 
probability rate that corresponds to that of Eq. (2.2).  
\vskip 0.2cm\noindent
(d) For each $b{\in}{\partial}{\Omega}$, we introduce two separate $V$\rq s, namely  
$h(b/L_{N})^{1/2}a_{b}^{\star}$ and $r_{b}^{1/2}a_{b}$, which lead to the creation 
and annihilation, respectively, of the particle at $b$, with weights corresponding to those 
of Eq. (2.2).
\vskip 0.2cm\noindent
Thus, under these specifications, the formula (2.13) takes the following form.
$${\cal G}A={\sum}_{x,y{\in}{\Omega}_{N}}^{\prime}
\bigl(a_{x}^{\star}a_{y}Aa_{y}^{\star}a_{x}-
{1\over 2}[a_{x}^{\star}a_{y}a_{y}^{\star}a_{x},A]_{+}\bigr)+$$
$$ {\sum}_{b{\in}{\partial}{\Omega}_{N}}
r_{b}\bigl(a_{b}^{\star}Aa_{b}-{1\over 2}[a_{b}^{\star}a_{b},A]_{+}\bigr)+ 
{\sum}_{b{\in}{\partial}{\Omega}_{N}}h(b/L_{N})
\bigl(a_{b}Aa_{b}^{\star}-{1\over 2}[a_{b}a_{b}^{\star},A]_{+}\bigr).
\eqno(2.14)$$
\vskip 0.3cm
{\it The Quantum Zero Range Model.} Since no exclusion principle is operative for this 
model, we take its particles to be bosons. Thus, the operators $a_{x}$ and 
$a_{x}^{\star}$ are unbounded here. In order to keep  the formulation of the model in 
terms, exclusively,  of bounded ones, we introduce the operators
$${\alpha}_{x}=(I+{\hat n}_{x})^{-1/2}a_{x}, \ {\alpha}_{x}^{\star}=
a_{x}^{\star}(I+{\hat n}_{x})^{-1/2}\eqno(2.15)$$
and note that, by Eqs. (2.6), (2.7) and (2.15), their actions on ${\psi}(n)$ are given by the 
formula
$${\alpha}_{x}{\psi}(n)=(1-{\delta}_{n_{x},0}){\psi}(n^{x,-}); \ 
{\alpha}_{x}^{\star}{\psi}(n)={\psi}(n^{x,+}).\eqno(2.16)$$
Hence, ${\alpha}_{x}^{\star}$ and ${\alpha}_{x}$ serve as bounded creation and 
annihilation operators. 
\vskip 0.2cm
In order to formulate the quantum version of the generator of the dynamical semigroup of 
the model, we proceed along the same lines as for the simple exclusion model, simply 
replacing $a_{x}$ by ${\alpha}_{x}$. Thus we obtain the following formula for the 
quantum version of Eq. (2.3), as applied to bosons.
$${\cal G}A={\sum}_{x,y{\in}{\Omega}_{N}}^{\prime}
\bigl(g({\hat n}_{x})^{1/2}{\alpha}_{x}^{\star}{\alpha}_{y}A{\alpha}_{y}
{\alpha}_{x}^{\star}g({\hat n}_{x})^{1/2}-
{1\over 2}[g({\hat n}_{x})^{1/2}{\alpha}_{x}^{\star}{\alpha}_{y}{\alpha}_{y}
{\alpha}_{x}^{\star}g({\hat n}_{x}^{1/2}),A]_{+}\bigr)+$$
$${\sum}_{b{\in}{\partial}{\Omega}_{N}}r_{b}\bigl(g({\hat n}_{b})^{1/2}
{\alpha}_{b}^{\star}A{\alpha}_{b}g({\hat n}_{b})^{1/2}-
{1\over 2}[g({\hat n}_{b})^{1/2}{\alpha}_{b}^{\star}{\alpha}_{b}
g({\hat n}_{b})^{1/2},A]_{+}\bigr)+$$
$${\sum}_{b{\in}{\partial}{\Omega}_{N}}h(b/L_{N})\bigl({\alpha}_{b}
A{\alpha}_{b}^{\star}-{1\over 2}[{\alpha}_{b}{\alpha}_{b}^{\star},A]_{+}\bigr).
\eqno(2.17)$$
\vskip 0.5cm 
\centerline {\bf 2.3. The Quantum System ${\Sigma}$ as an Extension of the Classical 
one ${\Sigma}_{cl}$}
\vskip 0.2cm
Since the subalgebra ${\cal B}$ of ${\cal A}$ is identified with that of the observables of 
${\Sigma}_{cl}$, the following proposition establishes that, for the simple exclusion and 
zero range models, the dynamics of ${\Sigma}$ induces an autonomous subdynamics on 
the classical observables ${\cal B}$ that is precisely that of the system ${\Sigma}_{cl}$. 
In other words the quantum system ${\Sigma}$ is an extension of the classical one, 
${\Sigma}_{cl}$.
\vskip 0.3cm
{\bf Proposition 2.2.} {\it For the models under consideration, 
\vskip 0.2cm\noindent
(1) The algebra ${\cal B}$ is stable under the semigroup 
${\phi}({\bf R}_{+})$; and
\vskip 0.2cm\noindent
(2) the restriction of ${\phi}({\bf R}_{+})$ to ${\cal B}$ is just the dynamical 
semigroup ${\phi}_{cl}({\bf R}_{+})$ of ${\Sigma}_{cl}$.}
\vskip 0.3cm
{\bf Proof .} Since ${\cal G}$ and ${\cal G}_{cl}$ are the generators of 
${\phi}({\bf R}_{+})$ and ${\phi}_{cl}({\bf R}_{+})$, respectively, it suffices to show 
that , for the models concerned, ${\cal G}_{cl}$ is just the restriction of ${\cal G}$ to 
${\cal B}$. By Prop. 2.1, this condition is just that      
$$[{\cal G}F({\hat n})]{\psi}(n)=[{\cal G}_{cl}F(n)]{\psi}(n) \ {\forall} \  
F{\in}{\cal C}({\Gamma}), \ n{\in}{\Gamma}.\eqno(2.18)$$
It is now a straightforward matter to check that, by Eqs. (2.5)-(2.8), (2.12), (2.14), (2.16) 
and (2.17), this condition is satisfied by both the simple exclusion and zero range models. 
\vskip 0.3cm
{\bf Proposition 2.3.} {\it  The dynamical transformations ${\phi}({\bf R}_{+})$ of the 
models under consideration commute with the gauge automorphisms 
${\gamma}({\theta})$. Hence the dynamics of these models are locally gauge covariant.}
\vskip 0.3cm
{\bf Proof.} Since ${\cal G}$ is the generator of ${\phi}({\bf R}_{+})$, it suffices to 
show that  ${\cal G}$ commutes with ${\gamma}({\theta})$; and it follows from Eqs. 
(2.4)-(2.6), (2.10), (2.11), (2.15) and (2.17) that it does so.
\vskip 0.5cm
\centerline {\bf  2.4. Steady States of ${\Sigma}_{cl}$ and ${\Sigma}$} 
\vskip 0.3cm
Assume now that ${\Sigma}_{cl}$ has a unique steady state ${\omega}_{cl}$, as has 
been established for the simple exclusion and zero range models [1-5]. We shall now 
show that ${\omega}_{cl}$ extends to a locally gauge invariant stationary state 
${\omega}$ of ${\Sigma}$. To this end we introduce the conditional expectation, $P$, of 
${\cal A}$ onto ${\cal B}$ that defines $PA$ as the mean over all the local gauge 
transformations ${\gamma}({\theta})$, i.e.
$$PA=
\Bigl[{\Pi}_{x{\in}{\Omega}_{N}}(2{\pi})^{-1}\int_{0}^{2{\pi}}d{\theta}_{x}\Bigr]
{\gamma}({\theta})A \ {\forall} \ A \ {\in}{\cal A}.\eqno(2.19)$$
We then define ${\omega}$ to be the state of ${\Sigma}$ given by the formula
$${\omega}(A)={\omega}_{cl}(PA) \ {\forall} \ A{\in}{\cal A}.\eqno(2.20)$$
In view of Props. 2.2 and 2.3, it follows immediately from the last two equations that 
${\omega}$ is indeed a locally gauge invariant stationary state of ${\Sigma}$, and that it 
is the only one that reduces to ${\omega}_{cl}$ on ${\cal B}$. Moreover, in the case of 
the models under consideration, it is the only stationary state\footnote*{In fact, the theory 
that follows does not depend on this uniqueness.} of ${\Sigma}$, for the following 
reasons. Frigerio [12, Theorem 3.2] has shown that a quantum dynamical semigroup 
${\phi}({\bf R}_{+})$ cannot admit more than one stationary state if the commutant of 
the operators $H$ and ${\lbrace}V_{j}{\rbrace}$ appearing in the formula (2.13) for its 
generator consists of the scalar multiples of the identity; and it follows easily from Eqs. 
(2.14) and (2.17) that, in view of the assumed strict positivity of the functions $h$ and 
$g$,  this condition is satisfied by the models under consideration. We remark that the 
steady states ${\omega}$ and ${\omega}_{cl}$ are ones of equilibrium or 
nonequilibrium according to whether or not the function $h$ is constant over 
${\partial}{\Omega}_{N}$. 
\vskip 0.2cm
In all cases, it follows from the above considerations that  the quantum dynamical system 
${\Sigma}$, as represented now by $({\cal A},{\phi},{\omega})$, is an extension of the 
classical one $({\cal B},{\phi}_{cl},{\omega}_{cl})$. 
\vskip 0.2cm
Since we shall be concerned with properties of the model in certain limits where $N$ 
tends to infinity, we shall henceforth indicate the $N$-dependence of ${\Sigma}, \ 
{\Sigma}_{cl}, {\omega}, \ {\omega}_{cl},  \ {\phi}, {\phi}_{cl}, \ {\cal G}$ and ${\cal 
G}_{cl}$ by attaching the superscript $(N)$ to these symbols. 
\vskip 0.5cm 
\centerline {\bf  3. The Hydrodynamic Picture.} 
\vskip 0.3cm 
We shall now investigate the large scale dynamical properties of the field ${\hat n}$, 
with the  aim of showing that it exhibits good hydrodynamical behaviour. To this end, we 
make the following two observations.
\vskip 0.2cm\noindent
(a) Since the field ${\hat n}$ of ${\Sigma}^{(N)}$ is built from the observables 
observables ${\hat n}_{x}$, which are affiliated to the algebra ${\cal B}$, it follows 
from Props. 2.1 and 2.2 that the dynamics of this field reduces to that of its classical 
counterpart, $n$ (of Section 2.1), as governed by the dynamical semigroup 
${\phi}_{cl}^{(N)}({\bf R}_{+})$ of ${\Sigma}_{cl}^{(N)}$.
\vskip 0.2cm\noindent
(b) As remarked in the Note following Eq. (2.3), the evolution of the field $n$ is 
diffusive. Hence, for a hydrodynamical description of this field on a length scale whose 
unit is $L_{N}$, the natural unit of the corresponding time scale is $L_{N}^{2}$.
\vskip 0.2cm
In view of these observations, we formulate the hydrodynamical picture 
of this field on macroscopic length and time scales whose units are $L_{N}$ and 
$L_{N}^{2}$, respectively, as in Refs. [1-5, 7, 8]. Thus, in this scaling, the dynamics of 
$n$ is represented formally by the classical field  
$$q_{t}^{(N)}(x)=
{\sum}_{y{\in}{\Omega}_{N}}{\phi}^{(N)}(L_{N}^{2}t)n_{y}{\delta}(y-L_{N}x) \ 
{\forall} \ x{\in}{\Omega}. \ y{\in}{\bf R}_{+}.$$
$q_{t}^{(N)}$ is therefore a ${\cal D}^{\prime}({\Omega})$-class distribution, in the 
sense of L. Schwartz [13]. To be precise, it is a continuous linear functional on the 
Schwartz space ${\cal D}({\Omega})$ of infinitely differentiable functions on ${\bf  
R}^{d}$ with support in ${\Omega}$; and its action on the latter space is given by the 
formula
$$q_{t}^{(N)}(f)=L_{N}^{-d}{\sum}_{y{\in}{\Omega}_{N}}
{\phi}_{cl}^{(N)}(L_{N}^{2}t)n_{y}f(L_{N}^{-1}y) \ {\forall} \ f{\in}
{\cal D}({\Omega}).\eqno(3.1)$$ 
\vskip 0.2cm
In fact this field $q_{t}^{(N)}$ does indeed exhibit good hydrodynamical properties  
since (cf. [2]), for  appropriate initial states ${\mu}^{(N)}$ of ${\Sigma}_{cl}^{(N)}$, 
the expectation value of $q_{t}^{(N)}(f)$ converges in probability, as 
$N{\rightarrow}{\infty}$, to the smeared form $\int_{\Omega}dxq_{t}(x)f(x)$ of a 
smooth field $q_{t}$, which evolves according to a phenomenological equation of the 
form
$${{\partial}q_{t}\over {\partial}t}={\Delta}{\Phi}(q_{t}), \eqno(3.2)$$
where the function ${\Phi}$ is smooth and non-negative: in the case of the simple 
exclusion model it is the identity function. The spatial boundary condition for this 
evolution is given by the formula
$${\Phi}\bigl(q_{t}(x)\bigr)=h(x) \ {\forall} \ x{\in}{\partial}{\Omega}, \ t{\in}
{\bf R}_{+},\eqno(3.3)$$ 
where $h$ is the function that governs the boundary term in Eqs. (2.2) and (2.3).
We denote by ${\overline q}$ the stationary solution of Eqs. (3.2) and (3.3). Evidently it 
is just the expectation value of $q_{t}^{(N)}$ for the nonequilibrium steady state 
${\omega}_{cl}$ in the limit $N{\rightarrow}{\infty}$. 
\vskip 0.2cm
We note that it follows from Eq. (3.2) that the linearised equation of motion for a  small 
perturbations ${\delta}q_{t}$ of ${\overline q}$ takes the form 
$${d\over dt}{\delta}q_{t}={\cal L}{\delta}q_{t}\eqno(3.4)$$
where
$${\cal L}={\Delta}\bigl[{\Phi}^{\prime}\bigl({\overline q}(x))(.)\bigr)\bigr].
\eqno(3.5)$$
Hence, assuming that ${\cal L}$ is the generator of a one-parameter semigroup 
${\lbrace}T_{t}{\vert}t{\in}{\bf R}_{+}{\rbrace}$ of linear transformations of 
${\cal D}^{\prime}({\Omega})$, the solution of Eqs. (3.4) is simply 
$${\delta}q_{t}=T_{t-t_{0}}{\delta}q_{t_{0}} \ {\forall} t{\geq}t_{0}\eqno(3.6)$$
\vskip 0.5cm 
\centerline {\bf  4. The Hydrodynamic Fluctuation Process}
\vskip 0.3cm
We define the field ${\xi}_{t}^{(N)}$, which represent the fluctuations of 
$q_{t}^{(N)}$ about  its mean for the steady state ${\omega}_{cl}^{(N)}$, by the 
formula
$${\xi}_{t}^{(N)}(f)=N^{1/2}\bigl[q_{t}^{(N)}(f)-
{\omega}_{cl}^{(N)}\bigl(q_{t}^{(N)}(f)\bigr)\bigr] \ {\forall} \ f{\in}{\cal 
D}({\Omega}).
\eqno(4.1)$$ 
${\xi}^{(N)}$ is thus a classical stochastic process for the state ${\omega}_{cl}$, 
indexed by ${\bf R}{\times}{\cal D}({\Omega})$. Our aim now is to verify that it 
satisfies the following conditions, which were the hypotheses on which the 
macrostatistical theory of  [7,8] was based.
\vskip 0.3cm
{\it (0) The hydrodynamic limit hypothesis.} This asserts that the process ${\xi}^{(N)}$ 
converges in law to a stationary stochastic process ${\xi}$ as $N{\rightarrow}{\infty}$.
\vskip 0.3cm
{\it (1). The Regression Hypothesis.} This asserts that the dynamical law governing 
deviations of the hydrodynamical variable $q_{t}$ from its steady state value is the same 
whether they arise from spontaneous fluctuations or from weak external perturbations.
Thus, in view of the formula (3.6) for the perturbed hydrodynamics, the regression 
hypothesis is that
$$E({\xi}_{t}{\vert}{\xi}_{t_{0}})=T_{t-t_{0}}{\xi}_{t_{0}} \ {\forall} \ 
t{\geq}t_{0}.\eqno(4.2)$$
where $E(.{\vert}{\xi}_{t_{0}})$ denotes the conditional expectation, given 
${\xi}_{t_{0}}$. 
\vskip 0.3cm
{\it (2) The Chaoticity Hypothesis.} In order to specify this hypothesis, we introduce 
Nelson\rq s [14] forward time derivative of ${\xi}_{t}$, namely
$$D{\xi}_{t}={\rm lim}_{{\tau}{\downarrow}0}
{\tau}^{-1}E\bigl({\xi}_{t+{\tau}}-{\xi}_{t}{\vert}{\xi}_{t}\bigr);\eqno(4.3)$$ 
and we infer from Eqs. (4.2) and (4.3) that, since ${\cal L}$ is the generator of 
$T({\bf R}_{+})$,
$$D{\xi}_{t}={\cal L}{\xi}_{t}.\eqno(4.4)$$
By the definition (4.3), $D{\xi}_{t}$ is the instantaneous expectation value of the rate of 
change of ${\xi}_{t}$. Accordingly, we designate $\int_{s}^{t}duD{\xi}_{u}$ to be the 
{\it secular} part of the increment $({\xi}_{t}-{\xi}_{s})$ in ${\xi}_{.}$ over the time 
interval $[s,t]$.  Correspondingly, we designate the {\it stochastic} part of $({\xi}_{t}-
{\xi}_{s})$ to be the remaining part, $w_{t,s}$, of this increment; and, in view of Eq. 
(4.4), this takes the form
$$w_{t,s}={\xi}_{t}-{\xi}_{s}-\int_{s}^{t}du{\cal L}{\xi}_{u}.\eqno(4.5)$$
Thus $w$ is a process indexed by ${\bf R}_{+}^{2}{\times}{\Omega}$. Our chaoticity 
hypothesis, which is designed to represent the stochasticity of this process, is that it is 
Gaussian and that its space-time correlations are of zero range, corresponding to ones of 
finite range on the microscopic scale. Thus the hypothesis is that $w$ is Gaussian and 
that
$$E\bigl(w_{t,s}(f)w_{t^{\prime},s^{\prime}}(g)\bigr)=0 \ {\rm if} \ either \ 
[s,t]{\cap}[s^{\prime},t^{\prime}]={\emptyset} \ or \  
{\rm supp}(f){\cap}{\rm supp}(g)={\emptyset}.\eqno(4.6)$$
\vskip 0.2cm 
{\it (3) The Local Equilibrium Hypothesis.} We formulate the local properties of the 
process ${\xi}$ in terms of the transformation $f{\rightarrow}f_{x_{0},{\epsilon}}$ of  
${\cal D}({\Omega})$  defined by the formula
$$f_{x_{0},{\epsilon}}={\epsilon}^{-d/2}f\bigl({\epsilon}^{-1}(x-x_{0})\bigr).
\eqno(4.7)$$
This transformation corresponds to the spatial rescaling by the factor ${\epsilon}$ around 
the point $x_{0}$. Further, in thermal equilibrium, the static two-point function for 
${\xi}$ enjoy the properties [2, 3] 
$$E\bigl({\xi}(f){\xi}(g)\bigr)={\chi}({\overline q})\int_{\Omega}dxf(x)g(x),$$
where ${\chi}$ represents the compressibility of the system; and
$$E\bigl({\xi}({\cal L}^{\star}(f){\xi}(g)\bigr)=E\bigl({\xi}(f)
{\xi}({\cal L}^{\star}g)\bigr)={\chi}({\overline q}){\Phi}^{\prime}
({\overline q})\int_{\Omega}dx{\nabla}f(x).{\nabla}g(x),$$
where ${\cal L}^{\star}$ is the dual of ${\cal L}$. By Eq. (4.6), these last two equations 
are equivalent to the following ones.
$$E\bigl({\xi}(f_{x_{0},{\epsilon}}){\xi}(g_{x_{0},{\epsilon}})\bigr)=
{\chi}({\overline q})\int_{\Omega}dxf(x)g(x),$$
and
$${\epsilon}^{2}E\bigl({\xi}({\cal L}^{\star}f_{x_{0},{\epsilon}})
{\xi}(g_{x_{0},{\epsilon}})\bigr)
={\epsilon}^{2}E\bigl({\xi}(f_{x_{0},{\epsilon}})
{\xi}({\cal L}^{\star}g_{x_{0},{\epsilon}})\bigr)={\chi}({\overline q}){\Phi}^{\prime}
({\overline q})\int_{\Omega}dx{\nabla}f(x).{\nabla}g(x).$$
The local equilibrium conditions, for fluctuations about nonequilibrium steady states, are 
just the limiting forms of these equations, as ${\epsilon}$ decreases to zero, with 
${\overline q}$ replaced by ${\overline q}(x_{0})$. Thus they are given by the formulae
$${\rm lim}_{{\epsilon}{\downarrow}0}
E\bigl({\xi}(f_{x_{0},{\epsilon}}){\xi}(g_{x_{0},{\epsilon}})\bigr)=
{\chi}\bigl({\overline q}(x_{0})\bigr) \int_{\Omega}dxf(x)g(x)\eqno(4.8)$$
and
$${\rm lim}_{{\epsilon}{\downarrow}0}{\epsilon}^{2}
E\bigl({\xi}(f_{x_{0},{\epsilon}}){\xi}({\cal L}^{\star}g_{x_{0},{\epsilon}})\bigr)
={\rm lim}_{{\epsilon}{\downarrow}0}{\epsilon}^{2}
E\bigl({\xi}({\cal L}^{\star}f_{x_{0},{\epsilon}})
{\xi}(g_{x_{0},{\epsilon}})\bigr)=$$
$${\chi}\bigl({\overline q}(x_{0})\bigr){\Phi}^{\prime}
\bigl({\overline q}(x_{0})\bigr)\int_{\Omega}dx{\nabla}f(x).{\nabla}g(x).
\eqno(4.9)$$
{\it Note.} These conditions represent local equilibrium on the {\it hydrodynamic} scale 
and are thus different from those formulated on the microscopic scale in Refs. [1, 2].
\vskip 0.5cm
\centerline {\bf 5. Verification of the Hypotheses 0-3 }
\vskip 0.3cm 
We shall now show that the above hypotheses are verified by the simple exclusion and 
zero range models. 
\vskip 0.3cm
{\it The Simple Exclusion Model.} The fluctuation process for this model was worked 
out in detail by Spohn [3] for the case where ${\Omega}$ is the slab\footnote*{Of 
course, for $d>1$, the slab does not meet our condition that ${\Omega}$ be bounded. 
However, it is a straightforward matter to extend our treatment and results to that 
situation.}, $(0,1){\times}{\bf R}^{d-1}$. Here we shall confine our attention to the one-
dimensional case, where ${\Omega}$ is the linear segment $(0,1)$.    
\vskip 0.2cm
For this case the following results have been established [3].
\vskip 0.2cm\noindent
(i) The function ${\Phi}$ appearing in the phenomenological law (3.2) is just the identity 
function and correspondingly, by Eq. (3.5), the generator ${\cal L}$ is the Laplacian 
${\Delta}$, with Dirichlet boundary conditions.
\vskip 0.2cm\noindent
(ii) The process ${\xi}_{t}^{(N)}$ converges in law, as $N{\rightarrow}{\infty}$, to a 
limit ${\xi}_{t}$ that is governed by a Langevin  equation 
$$dx_{t}={\cal L}{\xi}_{t}dt+dw_{t},\eqno(5.1)$$
where $w_{t}$ is the Wiener process for which
$$E\bigl([w_{t}(f)-w_{s}(f)][w_{t^{\prime}}(g)-w_{s^{\prime}}(g)]\bigr)=$$
$$2\int_{{\Omega}}dx{\chi}\bigl({\overline q}(x)\bigr){\Phi}^{\prime}
\bigl({\overline q}(x)\bigr){\nabla}f(x).{\nabla}g(x)
{\vert}[s,t]^{\cap}[s^{\prime},t^{\prime}]{\vert} \ {\forall} \ f,g{\in}
{\cal D}({\Omega}), \ t,s({\leq}t),t^{\prime},s^{\prime}({\leq}t^{\prime}){\in}
{\bf R}_{+}\eqno(5.2)$$
and 
$$E\bigl([w_{t}-w_{s}]{\vert}{\xi}_{u}\bigr)=0 \ {\rm for} \ 
t{\geq}s{\geq}u..\eqno(5.3)$$
\vskip 0.2cm\noindent
(iii) The static two-point function takes the form
$$E\bigl({\xi}(f){\xi}(g)\bigr)=\int_{0}^{1}dx{\chi}\bigl(q(x)\bigr)f(x)g(x)+
[h(1)-h(0)]\int_{0}^{1}dxf(x){\Delta}^{-1}g(x),\eqno(5.4)$$
where
$${\chi}(q)=q(1-q)\eqno(5.5)$$
and $h$ is the function appearing in the boundary term of Eq. (2.2).
\vskip 0.2cm
The result (ii) immediately substantiates the hypotheses (0) and (2). Moreover, since 
${\cal L}$ is the generator of ${\cal L}({\bf R}_{+})$, it follows from Eq. (5.1) that
$${\xi}_{t}=T_{t-t_{0}}{\xi}_{t_{0}}+\int_{t_{0}}^{t}T_{t-u}dw(u) \ {\forall}  \ 
t{\geq}t_{0}\eqno(5.5)$$
and hence, by Eq. (5.3), that the hypothesis (1) is also fulfilled. Finally, the local 
equilibrium properties (4.8) and (4.9) are simple consequences of the formulae 
(4.7) and (5.4), which signifies that the model also satifies hypothesis (3).
\vskip 0.3cm
{\it The Zero Range Model.} A key property of this model is that its steady state takes 
the simple product form [15, 5]
$${\omega}_{cl}^{(N)}={\otimes}_{x{\in}{\Omega}_{N}}
m_{x,{\overline q}(x)},\eqno(5.6)$$
where $m_{x,{\overline q}(x)}$ is a probability measure on the functions of $n_{x}$ 
that depends on the value of the stationary field ${\overline q}$ at the site $x$. 
\vskip 0.2cm
It follows now from a straightforward adaptation of the argument\footnote*{For that 
argument, as applied to the present situation, ${\xi}_{t}$ lies in the Sobolev space 
${\cal H}_{-r}({\Omega}):={\lbrace}f:{\Omega}{\rightarrow}
{\bf R}{\vert}\int_{\Omega}dxf(1-{\Delta})^{-n}f<{\infty}{\rbrace}$ for sufficiently 
large $r({\in}{\bf N})$. This space is a subspace of ${\cal D}^{\prime}({\Omega})$. 
The operators ${\cal L}^{\star}$ and $\bigl[{\Phi}\bigl(q(x)\bigr)\bigr]^{1/2}{\nabla}$ 
on ${\cal H}_{-r}$ play the roles of those denoted, in Gothic script, by ${\cal A}$ and 
${\cal B}$, respectively, in [2].} of [2, Ch. 11] that, for this model too, the process 
${\xi}_{t}^{(N)}$ converges in law to a Gaussian process ${\xi}$, which is also 
represented by Eqs. (5.1)-(5.3) , though with ${\cal L}$ now given by Eq. (3.5). Hence, 
by the same argument as for the simple exclusion model, we see that this model satisfies 
the hypotheses (0), (1) and (2). Further, it follows from (0) that
$$E\bigl({\xi}(f){\xi}(g)\bigr)={\rm lim}_{N\to\infty}{\omega}_{cl}^{(N)}
\bigl({\xi}^{(N)}(f){\xi}^{(N)}(g)\bigr)$$
and hence, by Eqs. (4.7) and (5.6), that the conditions (4.8) and (4.9) are fulfilled, with 
${\chi}\bigl({\overline q}(x)\bigr)$ the variance of the particle number at the site 
$L_{N}x$ in the single particle state ${\mu}_{x,{\overline q}(x)}$.
\vskip 0.5cm
\centerline {\bf 6. Concluding Remarks}
\vskip 0.3cm
In this article we have constructed a quantum stochastic model that fulfills the hypotheses 
of our general macrostatistical picture [7,8] of nonequilibrium steady states. A rather 
unphysical feature of this model is that the density field $q_{t}^{(N)}$ is classical not 
only at the hydrodynamical level but also at the microscopic one. Thus the problem of 
constructing a quantum model whose classical properties emerge only at the 
hydrodynamical and thermodynamical levels remains a challenging and interesting one.
\vskip 0.5cm
{\bf Acknowledgment.} Part of this work was carried out at the Workshop on Complex 
Systems at the E.S.I., Vienna, in the Spring of 2006. It is a pleasure to thank Michael 
Loss, Lazlo Erdos and Eric Carlen, the organisers of that Workshop, for the stimulating 
atmosphere they created there. I should also like to thank Alberto Barchielli and Franco 
Fagnola for informing me about Frigerio\rq s uniqueness theorem and Daniel Dubin for a 
helpful remark about the theory of distributions. 
\vskip 0.5cm
\centerline {\bf References}
\vskip 0.3cm\noindent
[1] A. De Masi, N. Ianiro, A. Pellegrinotti and E. Presutti: {\it A survey of the 
hydrodynamical properties of many-particle systems}, Studies in Statistical Mechanics, 
Vol. 10.
\vskip 0.2cm\noindent
[2] C. Kipnis and  C. Landim: {\it Scaling limits of interacting particle systems}, 
Springer, Berlin, Heidelberg, 1999.
\vskip 0.2cm\noindent
[3] H. Spohn: J. Phys. A. {\bf 16}, 4275-91, 1983.
\vskip 0.2cm\noindent
[4] B. Derrida, J. L. Lebowitz and E. R. Speer: J. Stat. Phys. {\bf 107}, 599, 2002. 
\vskip 0.2cm\noindent
[5] L.Bertini, A. Da Sole, D. Gabrielli, G. Jona-Lasinio and C. Landim: J. Stat. Phys. {\bf 
107}, 635-675, 2002 .
\vskip 0.2cm\noindent
[6] L. Onsager and S. Machlup: Phys. Rev. {\bf 91}, 1505, 1953.
\vskip 0.2cm\noindent
[7] G. L. Sewell: Lett. Math. Phys. {\bf 68}, 53, 2004.
\vskip 0.2cm\noindent 
[8] G. L. Sewell: Rev. Math. Phys. {\bf 17}, 977, 2005.
\vskip 0.2cm\noindent
[9] L.Onsager: Phys. Rev. {\bf 37}, 405, 1931.
\vskip 0.2cm\noindent
[10] V. Gorini, A. Kossakowski and E. C. G. Sudarshan: J. Math. Phys. {\bf  17}, 821, 
1976.
\vskip 0.2cm\noindent
[11]  G. Lindblad: Commun. Math. Phys. {\bf 48}, 119, 1976.
\vskip 0.2cm\noindent
[12] A. Frigerio:Commun. Math. Phys. {\bf 63}, 269, 1978.
\vskip 0.2cm\noindent
[13] E. Nelson: Ann. Math. {\bf 70}, 572, 1959.
\vskip 0.2cm\noindent
 [14]  L.Schwartz: {\it Theorie des Distributions}, Hermann, Paris, 1998.
\vskip 0.2cm\noindent
[15] A. De Masi and P. Ferrari: J. Stat. Phys. {\bf 36}, 81, 1984.
\end